\documentclass[twocolumn,showpacs,preprintnumbers,prb]{revtex4}
\newcommand{\BE}{\begin{equation}}
\newcommand{\EE}{\end{equation}}
\newcommand{\BA}{\begin{eqnarray}}
\newcommand{\EA}{\end{eqnarray}}

\usepackage{graphicx}% Include figure files
\usepackage{dcolumn}% Align table columns on decimal point
\usepackage{bm}% bold math
\input epsf.tex
\begin{document}

\title{Focusing of electromagnetic waves by
periodic arrays of dielectric cylinders}

\author{Bikash C. Gupta}
\author{Zhen Ye}\email{zhen@phy.ncu.edu.tw} \affiliation{Wave Phenomena
Laboratory, Department of Physics, National Central University,
Chungli, Taiwan 32054, R. O. C.}

%\date{February 2, 2002}
\date{\today}

\begin{abstract}

By numerical simulations, we show that properly arranged two
dimensional periodic arrays, formed by dielectric cylinders
embedded in parallel in a uniform medium, can indeed act as an
optical lens to focus electromagnetic waves, in accordance with
the recent conjecture in the literature. The numerical simulations
are based on an exact multiple scattering technique. The results
suggest that the E-polarized waves are easier to be focused than
the H-polarized waves. The robustness of the focusing against
disorders is also studied. Comparison with the corresponding cases
for acoustic waves is also discussed.

\end{abstract}

\pacs{42.70.Qs, 41.20.Jb, 42.25Lc} \maketitle

Photonic crystals (PCs) \cite{review,rev2} are made of
periodically modulated dielectric materials, and most sonic
crystals \cite{sigal} (SC) are made up of materials with periodic
variation of material compositions. Photonic and sonic crystals
have been studied both intensively and extensively. When passing
through photonic crystals, the propagation of electromagnetic (EM)
waves can be significantly affected by the photonic crystals in
the same way as that electrons are controlled by the crystals.
This interesting phenomenon has stimulated a variety of possible
applications of PCs in controlling photons. In particular,
considerable efforts have been devoted to finding photonic
crystals that can completely block propagation of electromagnetic
waves in all directions within a certain range of frequencies,
termed as photonic bandgap. It has been suggested that PCs may be
useful for various applications such as antennae \cite{brown},
optical filters \cite{chen}, lasers \cite{evans}, prisms
\cite{lin1}, high-Q resonant cavities \cite{lin2}, wave-guides
\cite{kraus1}, mirrors \cite{kraus2}, left-handed
materials\cite{Notomi,LHM1,LHM2}, and second harmonic generations
\cite{mart}. These applications mostly rely on the existence of
photonic bandgaps, and a majority of them is not concerned with
the linear dispersion region well below the first gap. In other
words, most of earlier studies were focused on the formation of
bandgaps and the inhibited propagation of waves.

Recently, the interest in the low frequency region, where the
dispersion relation is linear, has just started. Since the
wavelength in this region is very large compared to the lattice
constant, the wave sees the medium as if it were homogeneous, in
analogy with wave propagation in normal media. Consequently, a
possible new application of PCs has been suggested by a number of
authors\cite{Halevi1,Halevi2}. These authors suggested that PCs
could also be employed as custom-made optical components in the
linear regime below the first bandgap\cite{Halevi2}. However, no
physical realization of optical lenses has been made so far. Along
the same line of thought, it was suggested that SCs may also be
used to build acoustic lenses to converge the acoustic waves. A
necessary condition to be satisfied for constructing an acoustic
lens is that the acoustic impedance contrast between the SC and
the air should not be large; otherwise acoustic waves will be
mostly reflected. The recent experiment\cite{cervera} and the
corresponding numerical simulation\cite{Bikash} on acoustic waves
propagation through a lenticularly shaped SC have confirmed that
acoustic lenses by SCs are indeed possible. Encouraged by these
findings, in this paper we would like to further explore the
possibility of using PCs as an optical lens to focus
electromagnetic waves, following the line of the simulation of
acoustic lenses\cite{Bikash}.

In this paper we carry out numerical simulations on the focusing
of EM waves by PCs. We wish to theoretically realize the
particular predictions made in \cite{Halevi1,Halevi2}. Since the
multiple scattering technique has been successfully applied
earlier \cite{ye} to reproduce some experimental results on
acoustic propagation and scattering in SCs and this technique can
be fully adopted to EM waves, we will use this technique to study
the focusing effect of EM waves by PCs in detail. To the best of
our knowledge, there has been no earlier attempt in using the
multiple scattering theory to investigate the focusing phenomenon
of PCs.

The system considered here is similar to what has been presented
in \cite{Halevi1,Halevi2}. Assume that $N$ uniform dielectric
cylinders of radius $a$ are placed in parallel in a uniform
medium, perpendicular to the $x-y$ plane. The arrangement can be
either random or regular. The scattering and propagation of EM
waves can be solved by using the exact formulation of
Twersky\cite{Twersky}. While the details can be found in
\cite{ye}, here we only brief the main procedures. A unit
pulsating line source transmitting monochromatic waves is placed
at a certain position. The scattered wave from each cylinder is a
response to the total incident wave, which is composed of the
direct contribution from the source and the multiply scattered
waves from each of the other cylinders. The response function of a
single cylinder is readily obtained in the form of the partial
waves by invoking the usual boundary conditions across the
cylinder surface. The total wave ($E$ or $H$ for the E- or H-
polarization respectively) at any space point is the sum of the
direct wave ($E_0$ or $H_0$) from the transmitting source and the
scattered wave from all the cylinders. The normalized field is
defined as $T \equiv E/E_0 \ \mbox{or} \ H/H_0$; thus the trivial
geometrical spreading effect is eliminated.

\begin{center}
\begin{figure}[hbt]
\vspace{10pt} \epsfxsize=3.25in\epsffile{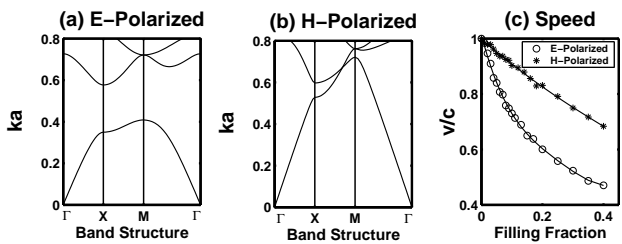} \caption{Band
structures for (a) the E- and (b) H- polarized electromagnetic
waves, and (c) the corresponding phase speeds for the two
dimensional square lattice of the dielectric cylinders in a
uniform medium.} \label{fig1}
\end{figure}
\end{center}

When the cylinders are placed regularly, the phenomenon of band
structures prevails, and can be evaluated by the standard
plane-wave expansion method. Fig.~\ref{fig1} shows the band
structures for both E- and H- polarized EM waves when propagating
through an array of square lattice of dielectric cylinders with
radius $a=0.38$ cm, placed in the air. The dielectric constant for
the cylinders is 10, which is smaller than that in \cite{Halevi1}.
The fractional area occupied by the cylinders for a unit area ,
i.~e. the filling factor, is 0.13. In the simulation, the
frequency is made non-dimensional by scaling as $ka$. Here is
shown that a complete bandgap appears for the E-polarized wave.
Following \cite{Halevi1,Halevi2}, we also calculate the phase
speed from the band structure for the first band as a function of
the filling factor, and the results are shown in
Fig.~\ref{fig1}~(c). It is the linear region of the first band
that we will consider in the following. The results indicate that
the phase speed for the E-wave is more significantly reduced by
the PC.

\begin{center}
\begin{figure}[hbt]
\epsfxsize=2.75in\epsffile{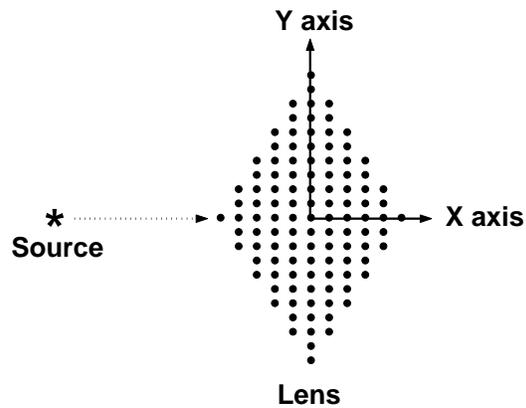} \caption{A lenticular
arrangement of the photonic crystal. The black filled circles
denote the dielectric cylinders. The coordinates used in the
simulation are shown in the figure.} \label{fig2}
\end{figure}
\end{center}

A PC made optical lens is illustrated in Fig.~\ref{fig2}, which is
in line with the prediction in \cite{Halevi1}. The source is
placed at a distance far enough from the lens so that the incident
waves can be regarded nearly as plane waves. In this way, the
focusing point of the lens can be inferred from the transmitted
field on the other side of the array.

In the rest computation, the following parameters and arrangements
are used: (1) Square lattices of cylinders and the propagation
along [10] direction are considered; (2) the dielectric constant
for the cylinders is 10; (3) the filling fraction is 0.13; (4) the
radius of cylinders is 0.38 cm; (5) the frequency of the waves is
taken as $ka=0.152$, well within the linear region; (6) the lens
thickness, i.~e. the max range along the $x$-axis, is 10, and the
height, i.~e. the max span along the $y$-axis, is 20; (7) the
distance between the source and the center of the lens  is 100.
All the lengths are scaled non-dimensionally by the lattice
constant.

\begin{center}
\begin{figure}[hbt]
\epsfxsize=2.75in\epsffile{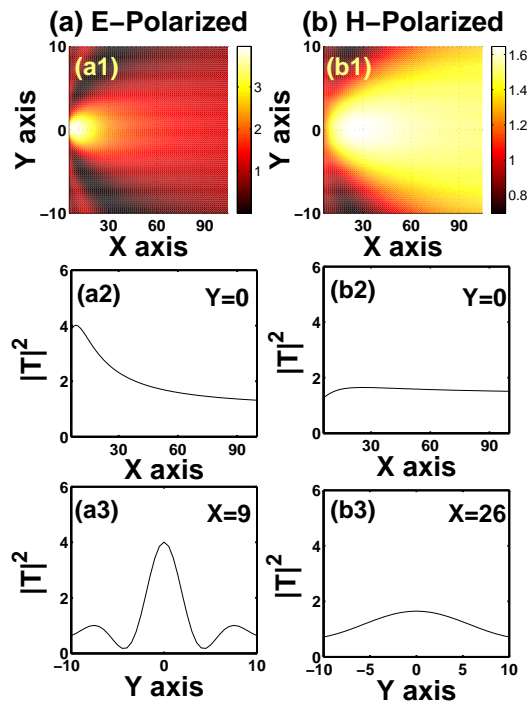} \caption{The two dimensional
spatial distribution of the transmitted intensity ($|T|^2$) on the
right side of the lens shown in Fig.~\ref{fig2}: (a) The
E-polarization, and (b) the H-polarization. (a2) and (b2): The
variation of the intensity along the $x$ axis at $y$=0. (a3) and
(b3): the variation of the intensity along the $y$ axis at $x=9$
and 26 respectively.} \label{fig3}
\end{figure}
\end{center}

Fig.~\ref{fig3} shows the two-dimensional spatial distributions of
the transmitted intensities. Here we see that although there are
some focusing effects for both E- and H- polarized waves, the
focusing effect is mostly prominent for the E-polarized wave in
particular. This agrees with the expectation from the phase speed
estimate in Fig.~\ref{fig1}~(c): the phase speed for the E-
polarized wave is more reduced, rendering a bigger contrast to the
outside medium, thereby yielding a bigger refractive index. The
focusing point of the E-polarized waves is at about $x=9$, and the
waves are better focused along the $y$-axis. From the results in
Fig.~\ref{fig3}, we see that both E- and H- polarized waves have
certain spreading along the x-axis, i.~e. the propagation
direction. This is very similar to what has been observed in the
acoustic lenses\cite{cervera,Bikash}. In any event, the fact that
the focusing features are in certain qualitative agreement with
the earlier prediction\cite{Halevi1,Halevi2} is encouraging.
Another note should be made here. In \cite{Halevi2}, Halevi et al.
conjectured an elliptically shaped lens using a 2D PC (See Fig.~1
in the paper). According to our earlier simulation on the acoustic
lenses, a lens of such a shape is less efficient than the
diamond-like lens illustrated by Fig.~\ref{fig2}.

\begin{center}
\begin{figure}[hbt]
\epsfxsize=2.75in\epsffile{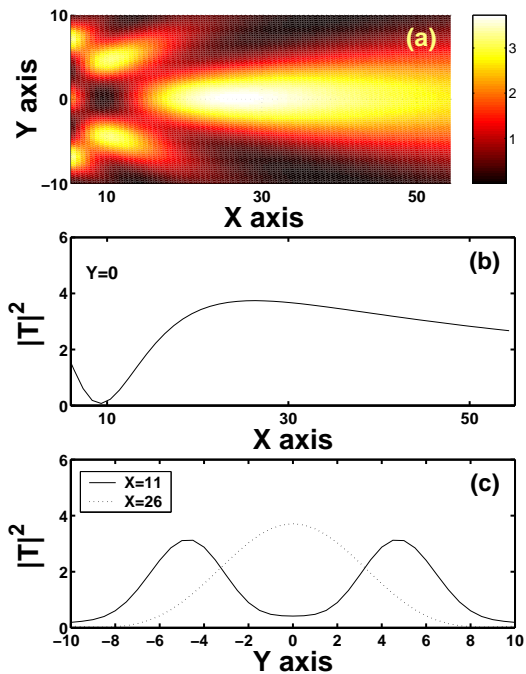} \caption{The two dimensional
spatial distribution of the transmitted intensity ($|T|^2$) on the
right side of a slab of photonic crystal for the E-polarization.}
\label{fig4}
\end{figure}
\end{center}

For comparison, we have also considered the E-polarized EM wave
transmission in the [10] direction through a slab of rectangular
array of dielectric cylinders. The size of the slab is $10\times
20$: the length along the $x$-axis is 10 and the height along the
$y$-axis is 20. The transmission results are shown in
Fig.~\ref{fig4}. Interestingly, there are some focusing effects.
For example, there is a focused field centered around $x=26, y=0$
and there are other two focusing centers at $x=11$ and $y=-5$ and
$5$ respectively. This is quite different from the corresponding
case with the acoustic arrays\cite{Bikash}. This seems not in the
expectation. The reason follows. As the source is quite far away
from the slab, the incident wave on the slab could be thought as a
plane wave. Then it is expected that the transmitted wave should
not be focused in the space when there is no such focusing effect
as shown in \cite{LHM2}. The results in Fig.~\ref{fig4} together
with that in Fig.~\ref{fig3} imply that although the focusing is
quite a general feature of a lattice arrangement of dielectric
cylinders, the shape of the lens as shown in Fig.~2 may be
essential for a unique focusing.

\begin{center}
\begin{figure}%[hbt]
\epsfxsize=2.75in\epsffile{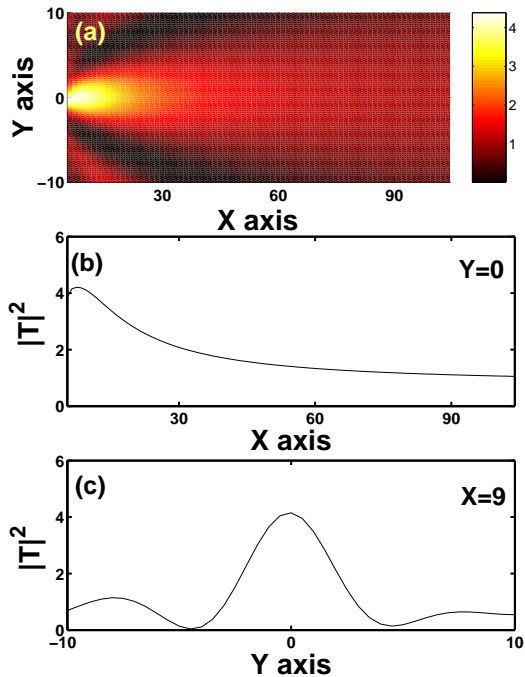} \caption{The two dimensional
spatial distribution of the transmitted intensity ($|T|^2$) after
passing one completely random configuration of the lattice; no
configuration averaging is taken. The shape of the sample is the
same as that in Fig.~\ref{fig2}.} \label{fig5}
\end{figure}
\end{center}

We have also examined the robustness of the focusing against
disorders. Here we consider the positional disorder of the
cylinders. Since the wavelength is larger than the lattice
constant in the present cases, one might intuitively conclude that
the positional disorder has no effects. The results for the
E-polarized wave are shown in Fig.~\ref{fig5}. Here the shape of
the array is the same as that in Fig.~\ref{fig2}, except that the
cylinders are placed in a complete randomness. Comparing to
Fig.~\ref{fig3}, it is shown that such a disorder does not destroy
the focusing, in accordance with the intuition. That the disorder
has little effect on the focusing of EM waves differs from the
situation with the acoustic systems. In \cite{Bikash}, it was
shown that the disorders can completely destroy the focusing
phenomenon in the acoustic lenses. All these results tend to
support the homogenization which has been carried out in
\cite{Halevi1}.

In summary, here we report the results of EM waves transmission
through lenticular structures made of dielectric cylinders.
Complying with the previous conjecture, the EM wave focusing
effects are indeed observable by such structures.

\acknowledgments{ This work received support from National Science
Council of Republic of China and the Department of Physics at the
National Central University.}

\end{document}